\documentclass[prd,twocolumn,nofootinbib]{revtex4}
\usepackage[utf8]{inputenc}

\pdfoutput=1
\usepackage{amsmath,amssymb,calc}
\usepackage{graphicx}
\usepackage{float}
\usepackage{amstext}    
\usepackage{array}    
\usepackage{color}
\usepackage{chngcntr}
\usepackage{comment}

\usepackage{cancel}
\usepackage{framed}
\usepackage{comment}
\usepackage{mathtools}
\usepackage{hyperref}
\usepackage{chngcntr}

\newcommand{\nn}{\nonumber}

   \def\e{\epsilon}
\def\d{\delta}
   \def\k{\kappa} 
 \def\o{\omega} \def\t{\tau} \def\l{\lambda}
\def\D{\Delta} \def\G{\Gamma} 
\def\L{\Lambda}

\newcommand{\Mcal}{{\mathcal M}} 
 \newcommand{\Ocal}{{\mathcal O}}

\begin{document}

\title{Global View of Axion Stars with (Nearly) Planck-Scale Decay Constants}
\def\Cincy{\small{Department of Physics, University of Cincinnati, Cincinnati, Ohio 45221,USA}}
\def\Weizmann{\small{Department of Particle Physics and Astrophysics,
Weizmann Institute of Science, Rehovot 7610001, Israel}}
\def\Harvard{\small{Department of Physics, Harvard University, 17 Oxford St., Cambridge, MA 02138, USA}}
\def\IPMU{\small{Kavli IPMU (WPI), UTIAS, The University of Tokyo, Kashiwa, Chiba 277-8583, Japan}}

\author{Joshua Eby}
\email[Electronic address:]{joshaeby@gmail.com}
\affiliation{\Weizmann}
\affiliation{\IPMU}

\author{Lauren Street}
\email[Electronic address:]{streetlg@mail.uc.edu}
\affiliation{\Cincy}

\author{Peter Suranyi}
\email[Electronic address:]{peter.suranyi@gmail.com}
\affiliation{\Cincy}

\author{L.C.R. Wijewardhana}
\email[Electronic address:]{rohana.wijewardhana@gmail.com}
\affiliation{\Cincy}

\begin{abstract}
 We show that axion stars formed from axions with nearly Planck-scale decay constants $f$ are unstable to decay, and are unlikely to have phenomenological consequences. More generally, we show how results at smaller $f$ cannot be naively extrapolated to $f=\Ocal(M_P)$ as, contrary to conventional wisdom, gravity and special relativity can both become relevant in the same regime. We clarify the rate of decay by reviewing and extending previous work on oscillons and axion stars, which imply a fast decay rate even for so-called dilute states at large $f$. 
\end{abstract}

\maketitle

\section{Introduction}

Axions \cite{Peccei:1977hh,Peccei:1977ur,Weinberg:1977ma,Wilczek:1977pj,Dine:1981rt,Zhitnitsky:1980tq,Kim:1979if,Shifman:1979if} and axion-like particles (ALPs) \cite{Turner:1983he,Press:1989id,Sin:1992bg,Hu:2000ke,Goodman:2000tg,Peebles:2000yy,Amendola:2005ad,Li:2013nal,Marsh:2015xka,Hui:2016ltb,Lee:2017qve} originate in numerous physical theories of physics beyond the Standard Model. Their low-energy phenomenology is governed by two energy scales: the particle mass $m$ and the decay constant $f$. If the axions are pseudo-Goldstone bosons, they can be described by a periodic potential that respects an approximate shift symmetry on the axion field $\phi$, commonly taking the form 
\begin{equation} \label{Vphi}
 V(\phi) = m^2 f^2\,\left[1 - \cos\left(\frac{\phi}{f}\right)\right].
\end{equation}
The leading self-interaction term in the expansion of $V(\phi)$ at $\phi\ll f$ gives rise to a $\lambda_4\phi^4$ potential with attractive coupling $\lambda_4 = -m^2/f^2$; higher-order self-interaction terms become relevant at high densities \cite{Eby:2016cnq,Visinelli:2017ooc}. Such fields might be probed by current and near-future experiments, even if they possess only gravitational couplings to ordinary matter \cite{Grin:2019mub}.

Light scalars (like axions and ALPs) can constitute dark matter (DM) in the universe \cite{Preskill:1982cy,Abbott:1982af,Dine:1982ah}, and field overdensities can collapse to form bound states known as boson stars (here, \emph{axion stars}); these are held together through a balance of kinetic pressure, gravitational attraction, and self-interactions \cite{Kaup:1968zz,Ruffini:1969qy,BREIT1984329,Colpi:1986ye,Seidel:1990jh,Friedberg:1986tq,Seidel:1991zh,Liddle:1993ha,Lee:1991ax,Chavanis:2011zi,Chavanis:2011zm,Eby:2016cnq}. The standard lore is that axion stars are solutions of the equations of motion falling into three distinct classes of increasing density, known respectively as the \emph{dilute}, \emph{transition}, and \emph{dense} branches of solutions, each with distinct macroscopic properties and stability constraints (see for example \cite{Visinelli:2017ooc,Braaten:2019knj,Braaten:2018nag,Eby:2019ntd}).

Axion stars on the dilute branch are generally stable both under perturbations \cite{Chavanis:2011zi,Chavanis:2011zm} and to decay \cite{Eby:2015hyx,Eby:2017azn}. Recently both simulations \cite{Schive:2014hza,Levkov:2018kau,Eggemeier:2019jsu} and analytic arguments \cite{Kirkpatrick:2020fwd} have suggested that they can efficiently form in the early universe; as such, dilute axion stars have been investigated for possible phenomenological effects, including recent analyses of radio photon emission \cite{Hertzberg:2018zte,Hertzberg:2020dbk,Levkov:2020txo,Amin:2020vja} and gravitational lensing \cite{Croon:2020wpr,Prabhu:2020pzm,Croon:2020ouk}.
For the attractive $\phi^4$ potential, there is a maximum stable mass of order $M_P/\sqrt{|\l_4|} = M_P\,f/m$ (for $\l_4 = -m^2/f^2<0$, and with $M_P=1.22\times10^{19}$ GeV) which signals a crossover to the structurally unstable transition branch \cite{Chavanis:2011zi,Chavanis:2011zm,Eby:2014fya,Schiappacasse:2017ham}. Because transition states are unstable to perturbations, they are unlikely to have observable consequences.

On the third branch, so-called dense axion stars have received considerable interest in recent years. The ``dense" moniker was coined in \cite{Braaten:2015eeu} when the configurations were investigated using the nonrelativistic equation of motion; note however that they are fundamentally no different from axitons, discussed decades earlier in a cosmological context by Kolb and Tkachev \cite{Kolb:1993hw}. Today, dense axion stars / axitons are understood to be strongly bound with momentum dispersion of order $m$, and as a consequence, they decay to relativistic axions with a short lifetime and are unlikely to have phenomenological consequences \cite{Eby:2016cnq,Visinelli:2017ooc}.

In this work, we investigate the nature of axion stars at large $f$, approaching the Planck scale. The reader may wonder whether the case of large $f$ is well-motivated enough to warrant a full investigation; to allay this critique, we review several contexts in which this parameter space is commonly invoked:

\begin{itemize}
 \item[1.] The QCD axion \cite{Peccei:1977hh,Peccei:1977ur,Weinberg:1977ma,Wilczek:1977pj,Dine:1981rt,Zhitnitsky:1980tq,Kim:1979if,Shifman:1979if} is motivated by its ability to solve the strong CP problem while simultaneously producing a good DM candidate. This model imposes a relation between the two relevant scales such that $m\,f \simeq 6\times10^{-3}$ GeV$^2$ \cite{DiVecchia:1980yfw,diCortona:2015ldu}. The abundance of dark matter QCD axions depends on cosmological assumptions, with masses $m \gtrsim \mu$eV required if the global symmetry of the axion is broken after inflation, corresponding to relatively low $f \lesssim 6\times10^{12}$ GeV (see \cite{Gorghetto:2020qws} for a recent, more stringent bound on the post-inflationary scenario). On the other hand, if symmetry breaking occurs before inflation, then masses all the way down to $m \simeq 5\times10^{-13}$ eV are allowed (at the expense of tuning the initial misalignment angle, see e.g. \cite{diCortona:2015ldu}, or non-minimal model-building, e.g. \cite{Bonnefoy:2018ibr}), in which case $f\simeq M_P$ is obtained. Even in such a model, where there are no large primordial density fluctuations, it may be possible to form axion stars through large misalignment \cite{Arvanitaki:2019rax}.
 
 \item[2.] In other particle physics models, the scales $m$ and $f$ characterizing the axion are independent.
ALPs may, for example, originate from string theory at low-energy from string compactification or as a consequence of anomaly cancellation \cite{Svrcek:2006yi,Cicoli:2012sz,Arvanitaki:2009fg}. In such cases the decay constant $f$ is naturally very large, of order the Grand Unified Theory (GUT) scale $\L_{GUT}\simeq 10^{16}$ GeV or higher. A widely-known example is Ultralight DM (ULDM) \cite{Hu:2000ke,Press:1989id,Sin:1992bg,Turner:1983he,Peebles:2000yy,Amendola:2005ad,Marsh:2015xka,Hui:2016ltb,Lee:2017qve}, where $m\simeq10^{-22}-10^{-20}$ eV implies galaxy-scale axion wavelengths, and axion stars forming in the central cores of galaxies \cite{Schive:2014dra,Schive:2014hza,Mocz:2017wlg,Veltmaat:2019hou,Nori:2020jzx}. Another well-motivated particle physics model is coherent relaxion dark matter, which may also require $f \gtrsim \L_{GUT}$ in order to be consistent with constraints from fifth-force experiments \cite{Banerjee:2018xmn}.

 \item[3.] A more phenomenologically-driven parameter choice is $f\sim M_P$ and $m\sim10^{-10}$ eV, as the dilute bound states become very compact, having (at their maximum stable mass) roughly the size and mass of neutron stars: $R=\Ocal$(few) km and $M=\Ocal(M_\odot)$. Such states may collide with other astrophysical bodies, giving rise to gravitational wave signals \cite{Clough:2018exo,Dietrich:2018jov}.

\item[4.] Axion stars have also been investigated as the origin of certain black holes in the universe.
At small $f$, it is known that axion stars which grow unstable and collapse do not form black holes, as they are stabilized by a core repulsion; they instead explode in a burst of scalar radiation in their approach to the dense configuration, a process known as a Bosenova \cite{Eby:2016cnq,Eby:2017xrr,Levkov:2016rkk,Helfer:2016ljl}. But at large $f\gtrsim0.1\,\tilde{M}_P$ (where $\tilde{M}_P = 2.4\times10^{18}$ GeV is the reduced Planck mass), it has been suggested that collapse can indeed lead to formation of black holes \cite{Helfer:2016ljl,Chavanis:2017loo,Michel:2018nzt,Widdicombe:2018oeo}. If true, this process could explain the tentative recent observation of a black hole of $M \sim 2-3M_\odot$ by the LIGO/VIRGO collaborations \cite{Abbott:2020khf}, which lies in a mass gap not easily explained by standard theories of black hole formation.
\end{itemize}
On the basis of this and other work, we conclude that the parameter range $\L_{\rm GUT} \lesssim f \lesssim M_P$ is motivated, and warrants the study we put forward here.

In this work, we show that the naive picture of three branches of axion stars outlined above breaks down as the decay constant $f$ approaches $M_P$, due to the usual assumptions about the relevance of gravity and special relativity breaking down. This fact implies that previous naive estimations of axion star parameters in this regime have neglected important contributions. 
In addition, we will point out that axion stars with large $f$ become unstable to decay to relativistic particles, even on the dilute branch;
as a result, such axion stars are unlikely to be phenomenologically relevant in the manner described above. 
We will make this point by reviewing calculations for the decay rate in previous literature, extending them to include gravity and higher-order self-interactions, and finally determining the full range of stable axion star solutions.

We will use natural units throughout, where $\hbar = c = 1$.

\section{Axion Stars}

\subsection{Non-Relativistic Bound States} \label{sec:RB}

We first review a few 
relevant facts about axion stars,
which are derivable using a number of viable methods \cite{Eby:2019ntd};
for our purposes, the formalism of Ruffini and Bonazzola (RB) \cite{Ruffini:1969qy} is the most useful.
RB used an expansion of the axion field operator which was linear in creation and annihilation operators to describe axionic bound states when self-interactions were absent; their work was first extended to the case of an attractive $\phi^4$ self-interaction in \cite{Barranco:2010ib}. Some of the present authors further extended the analysis to fully characterize the attractive $\phi^4$ case \cite{Eby:2014fya}, and also to include contributions from scattering states that give rise to decay processes \cite{Eby:2015hyx,Eby:2017azn} and relativistic corrections to bound states \cite{Eby:2017teq}.

The generic, spherically-symmetric RB field operator can be written in the form
\begin{align} \label{RB}
 \phi(t,r) &= R(r)\left[e^{-i\,\e\,m\,t}\,a_0 + h.c.\right]  &{\rm (RB)} \nn \\
 		&\qquad + \sum_{k>1}^{\infty} R_k(r)\left[e^{-i\,k\,\e\,m\,t}\,a_0^k + h.c.\right]  &{\rm (GRB)} \nn \\
		&\qquad + \left[\psi_f(t,r) + h.c.\right],  &{\rm (Scattering)}
\end{align}
where $R(r)$, $R_k(r)$, and $\psi_f(t,r)$ are the $N$-particle bound ground state, higher-harmonic states, and single-particle scattering state wavefunctions (respectively), $a_0$ and its conjugate are the annihilation and creation operators for the bound state, and $\e\,m<m$ is the bound state eigenenergy. 
The scattering and higher-order harmonic modes were not included by RB, and will be discussed in the next sections.
In the weak binding limit, where $|\e-1| \ll 1$, it is appropriate to rescale the wavefunction as $Y(x) = 2\sqrt{N}R(r)/(f\,\D)$ with the coordinate rescaled as $x = \D\,m\,r$ \cite{Eby:2014fya}, where $\D \equiv \sqrt{1 - \e^2}$.

\begin{table}[b]
\begin{tabular}{|c ||c |c |c |c | }
\hline
 Color & Red & Green & Brown & Purple
  \rule{0pt}{2.6ex} \rule[-1.2ex]{0pt}{0pt} \\ \hline

 $\d$ & $10^{-4}$ & $10^{-3}$ & $10^{-2}$ & $10^{-1}$ 
 \rule{0pt}{2.6ex} \rule[-1.2ex]{0pt}{0pt} \\ \hline
 
 $f$ [GeV] & $2.4\times 10^{16}$ & $7.7\times10^{16}$ & $2.4\times 10^{17}$ & $7.7\times10^{17}$ 
  \rule{0pt}{2.6ex} \rule[-1.2ex]{0pt}{0pt}\\  \hline
 
 $f/M_P$ & $2\times 10^{-3}$ & $6.3\times10^{-3}$ & $0.02$ & $0.063$ 
  \rule{0pt}{2.6ex} \rule[-1.2ex]{0pt}{0pt}\\  \hline
   
 $f/\tilde{M}_P$ & $0.01$ & $0.032$ & $0.1$ & $0.32$ 
  \rule{0pt}{2.6ex} \rule[-1.2ex]{0pt}{0pt}\\  \hline
 
 $f/\L_{\rm GUT}$ & $2.4$ & $7.7$ & $24$ & $77$ 
 \rule{0pt}{2.6ex} \rule[-1.2ex]{0pt}{0pt} \\  \hline
 
 $\k$ ($\D=0.1$) & $0.01$ & $0.1$ & $1$ & $10$ 
 \rule{0pt}{2.6ex} \rule[-1.2ex]{0pt}{0pt}\\ \hline
\end{tabular}
\caption{Relationship between the expansion parameter $\d$, the decay constant $f$, and the effective gravitational coupling $\k$, and the corresponding color in the Figures, over the range of inputs considered in this work. Note for reference that $M_P = 1.22\times10^{19}$ GeV, $\tilde{M}_P = 2.4\times10^{18}$ GeV and $\L_{\rm GUT} = 10^{16}$ GeV.}
 \label{tab:fdelta}
\end{table}

Bound-state configurations can be determined by solving the coupled Einstein+Klein-Gordon (EKG) equations for the wavefunction $Y(x)$, along with the functions determining the gravitational metric. In the Newtonian and nonrelativistic limits, the equations of motion are \cite{Eby:2014fya}
\begin{align}
 \nabla^2Y(x)&= -\frac{1}{8}\, Y(x)^3 
	  + [1 + \kappa\,b(x)]\,Y(x), \label{EKGY} \\ 
 b(x) &= -\frac{1}{8\pi}\left[\int d^3 x' 
	\frac{Y(x')^2}{|\vec{x}-\vec{x}'|}\right], \label{EKGb}
\end{align}
where the spherically-symmetric metric function\footnote{Note the appearance of $8\pi$ in definition of $\d$, which did not appear in \cite{Eby:2014fya}; this is because here we define the expansion in terms of $M_P=1.22\times10^{19}$ GeV rather than the reduced Planck mass $\tilde{M}_P = 2.4\times10^{18}$ GeV.}
$g^{tt} = 1 - \delta\,b(x)$ 
with $\delta = 8\pi\,f^2/M_P^2$; the other relevant metric function $g^{rr} = 1-\d\,a(x)$ was eliminated using the $rr$ Einstein equation. Note that we have taken the potential to be that of Eq. \eqref{Vphi};
it has been shown that in this case the above equations are equivalent to the Schr\"odinger+Poisson equations often used in the study of boson stars \cite{Eby:2018dat}.

The set of equations (\ref{EKGY}-\ref{EKGb}) represents the leading order in a double-expansion of the relativistic EKG equations in the two small parameters $\d \ll 1$ (representing weak, Newtonian gravity) and $\D \ll 1$ (representing weak binding, or the nonrelativistic limit); as written, they are correct to $\Ocal(\d,\D^2)$ \cite{Eby:2014fya}, and in the next section we extend to the next order in special-relativistic corrections, $\Ocal(\D^4)$. Previous perturbative studies have been restricted to either the fully nonrelativistic limit (e.g. \cite{Chavanis:2011zi} and many others), or they include leading-order special relativity but neglect Newtonian gravity \cite{Mukaida:2016hwd,Braaten:2016kzc,Namjoo:2017nia,Braaten:2018lmj}; in the next section, we attempt to bridge this gap by including both effects. For a detailed discussion of the relativistic expansion parameters, see \cite{Eby:2019ntd}.

There exist higher-order corrections proportional to $\d\,\D^2$ or $\d^2$, which we require to be subleading by restricting the maximum parameter values $\d_{\rm max} = 0.1 \ll \D_{\rm max}^2 = 1/2$ throughout; the effect of these subleading terms was partially investigated in \cite{Croon:2018ybs}, but for simplicity we leave a full study for future work.

The one free parameter in Eqs. (\ref{EKGY}-\ref{EKGb}) is $\k = \d/\D^2$, which is the effective coupling to gravity. For a given value of $\k$, these coupled equations have a unique solution that is easily determined by the standard shooting method: one varies the central values $Y(0)$ and $b(0)$ until there is exponential convergence $Y(x \to x_{\rm max}) \to 0$ and $a(x \to x_{\rm max}) + b(x \to x_{\rm max}) \to 0$ at some large $x_{\rm max}$. This can be done to arbitrarily high precision (see \cite{Eby:2014fya} for details). For clarity we translate between values of $\d$, $f$, and $\k$ in Table \ref{tab:fdelta} for the range we investigate here.

Eqs. (\ref{EKGY}-\ref{EKGb}) are, in most applications, sufficient to describe the dilute and transition branches of axion star solutions. On the stable dilute branch, the total mass $M \propto \D$ and the radius 
$R \propto \D^{-1}$, so that $M\propto R^{-1}$. Stable solutions have $\k \geq \k_c \equiv 0.34$; the inequality is saturated at a maximum stable mass $M_{\rm c} = 10.2\,M_P\,f/m$, which corresponds to a critical value of 
\begin{equation} \label{eq:Deltac}
\D_c \approx \sqrt{\frac{8\pi}{\k_c}}\,\frac{f}{M_P} = 8.6\,\frac{f}{M_P}.
\end{equation}
Beyond this critical point, the central density $Y(0)$ grows while the gravitational coupling $\k$ continues to decrease below unity, leading to a decoupling of gravity which characterizes the transition branch.

From here we can already see that naive extrapolation to $f=\Ocal(M_P)$ will fail, as Eq.~\eqref{eq:Deltac} predicts $\D_c = \Ocal(1)$ or larger. In reality, we will find in the next section that above some value of $f$ this naive maximum mass point is no longer attainable.
Indeed, when $\D$ becomes large, the eigenenergy in the axion star approaches $m$, implying a breakdown of the nonrelativistic approximation \cite{Eby:2019ntd}. This also affects the bound state and
leads to relativistic decay processes that render the star unstable; we discuss both effects below.

\subsection{Bound States with Relativistic Corrections} \label{sec:GRB}

There have recently been several independent efforts to quantify the effect of relativistic corrections on axion stars, which are relevant at large densities \cite{Mukaida:2016hwd,Braaten:2016kzc,Namjoo:2017nia,Braaten:2018lmj}. In one such work, it was shown how such corrections can be organized as a power series in the parameter $\D$ by extending the RB framework, thereby defining a Generalized Ruffini-Bonazzola (GRB) procedure \cite{Eby:2017teq}. In that work gravity was negligible, because (a) the focus was on the transition / dense branches of solutions, and (b) the input parameters were assumed to be in the standard range for QCD axion experiments, where $f \sim 10^{10-12}$ GeV. However at large $f$, the parameters $\delta$ and $\Delta$ can be of the same order, and therefore both Newtonian gravity and special-relativistic corrections must be taken into account.

At leading order in higher-harmonic (GRB) corrections of Eq.~\eqref{RB},
Eq.~\eqref{EKGY} is modified as \cite{Eby:2017teq}
\begin{align} \label{eq:GRB}
 \nabla^2Y(x)&= -\frac{1}{8}\, Y(x)^3 + [1 + \kappa\,b(x)]\,Y(x) + \frac{3}{512}\Delta^2\,Y(x)^5,
\end{align}
where we continue to use the potential of Eq. \eqref{Vphi}. Once again, note that we ignore corrections proportional to $\d\,\D^2$, which appeared in \cite{Croon:2018ybs}, or proportional to $\d^2$, which are post-Newtonian, because such corrections are small for the parameter space we consider.

The mass $M$ of the axion star at leading-order in GRB is $M=\Mcal(x_{\rm max}\to \infty)$ with\footnote{For more information, see \cite{Eby:2017teq} as well as Appendix B of \cite{Eby:2017zyx}.}
\begin{align} \label{eq:mass}
 \Mcal(x_{\rm max}) &= \frac{\pi\,f^2}{m\,\D}\int_0^{x_{\rm max}}dx\,x^2\Big[(2-\D^2+\d\,a(x))Y(x)^2 \nn\\
 			&\quad + \D^2\,Y'(x)^2 -\frac{\D^2}{16}Y(x)^4 + \frac{\D^4}{512}Y(x)^6\Big].
\end{align}
We define the radius as $R_{99}$, inside which $0.99$ of the mass is contained,
\begin{equation} \label{eq:R99}
 \Mcal(\D\,m\,R_{99}) = 0.99\times M.
\end{equation}
To the extent that the number of particles is conserved---a topic we will return to in the next section---we can treat axion stars as $N$-particle states with $N=M/m$.

\begin{figure}[t]
 \includegraphics[scale=.22]{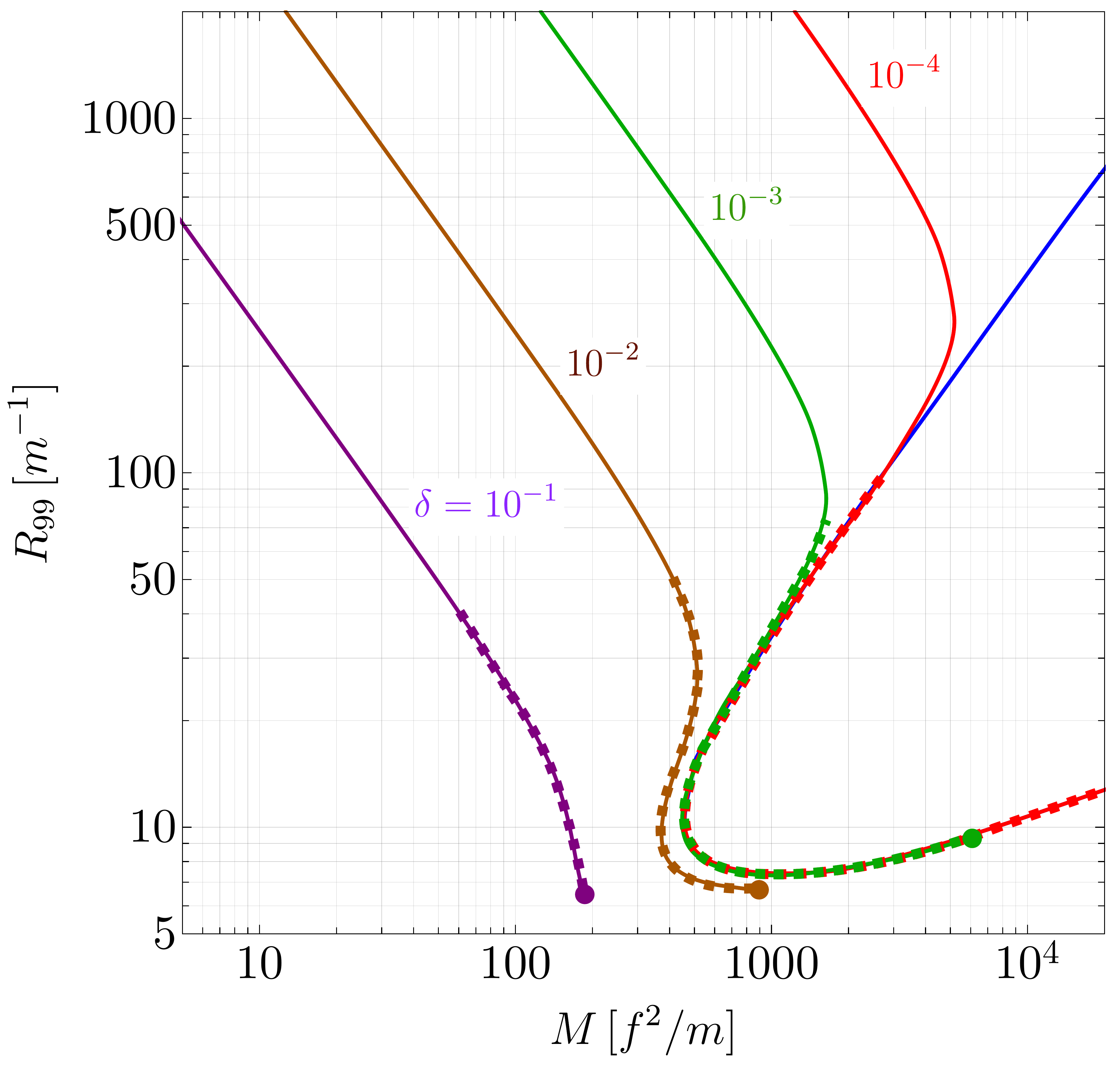}
 \caption{The mass-radius relation for axion stars at leading-order in GRB, obtained by solving Eqs. (\ref{EKGb}\,,\,\ref{eq:GRB}-\ref{eq:R99}). The curves are labeled by values of $\d$ and color-coded as in Table \ref{tab:fdelta}; the blue curve corresponds to the limit of gravity decoupling, $\k\to0$. Configurations on the solid curves are stable to decay processes with long lifetimes $\tau_0 > \tau_U$; those along the dashed curves are unstable to decay. The filled circles represent the endpoint of each curve at $\D = 1/\sqrt{2}$.}
 \label{fig:MvsR}
\end{figure}

\begin{figure}[t]
 \includegraphics[scale=.60]{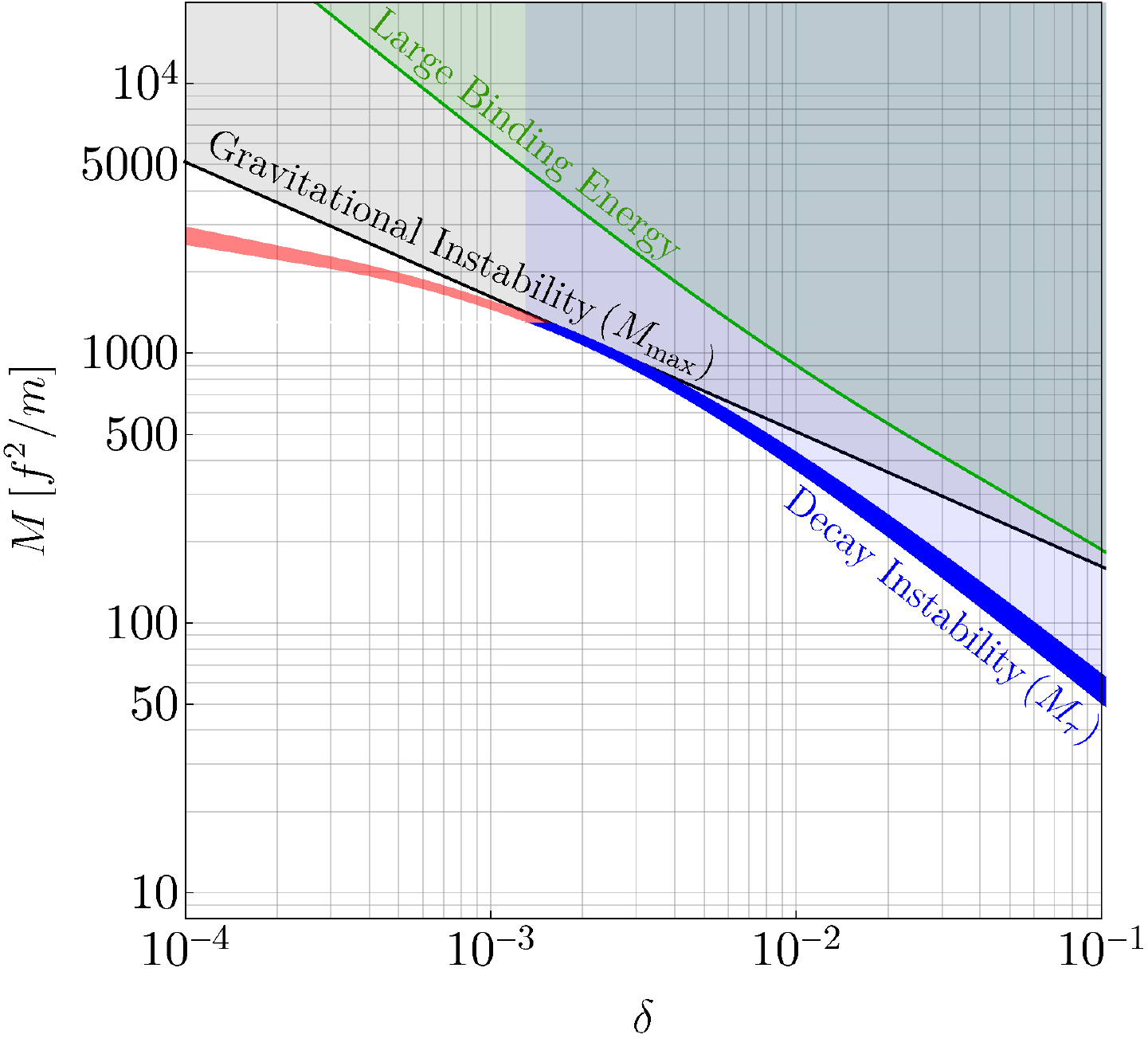}
 \caption{Critical lines of stability for axion stars in the plane of mass $M$ vs $\delta$. The black shaded region marks gravitational instability, $M>M_{\rm max}$; the green shaded region marks large binding energies $\D > 1/\sqrt{2}$; and the blue shaded region denotes parameter space on the dilute branch of solutions with short lifetimes $\tau_0 < \tau_U$. The red band denotes the crossover to decay instability on the transition branch. For both blue and red bands, the width represents the axion particle mass varied in the range $10^{-20}$ eV $\leq m \leq 1$ eV.}
 \label{fig:Mcrit}
\end{figure}

The system of equations (\ref{EKGb}\,,\,\ref{eq:GRB}) requires two input parameters, $\Delta$ and $\kappa$, though at fixed $\D$ we can trade $\k$ for the more intuitive parameter
$\d$. Then the solutions have masses and radii computable using Eqs. (\ref{eq:mass}-\ref{eq:R99}); the results are depicted in Fig. \ref{fig:MvsR}, where the curves are labeled by values of $\d$ and color-coded as in Table \ref{tab:fdelta}. The blue line corresponds to $\k\to 0$ or full gravitational decoupling. The endpoints of each curve (marked with a filled circle) represent 
$\D=\D_{\rm max} = 1/\sqrt{2}$, essentially an arbitrary cutoff; at $\D_{\rm max}$, not only does the GRB expansion break down, but the binding energy is so large that the momentum uncertainty becomes comparable to the mass $m$, and the usual methods of analysis fail \cite{Eby:2019ntd}. Note for future reference that on the red curve, $f \simeq 2.4\,\L_{\rm GUT}$, and on the purple curve $f \simeq 0.3\,\tilde{M}_P$.

For low $f\lesssim 10^{17}$ GeV, we see in Fig. \ref{fig:MvsR} the standard separation into three distinct branches of solutions. The dilute and transition branches are joined at the (local) maximum mass $M_{\rm max}$, while the transition and dense branches are joined at a (local) minimum mass $M_{\rm min} \sim {\rm few}\times100 f^2/m$. As $f$ grows, we find at this order in $\D$ that the transition branch shrinks, while simultaneously the cutoff at high $\D$ limits the extent of the dense branch, leaving only the dilute branch intact. It is possible that between our cutoff $\D_{\rm max} = 1/\sqrt{2}$ and the maximum possible value $\D=1$ that some portion of the dense branch persists; however, the transition branch is almost certainly lost, as we see the dilute and dense branches joining to a single point between $\d=10^{-1}$ and $\d=10^{-2}$.

In previous literature, it has always been assumed that the region where relativistic corrections are important is nonoverlapping with the region where gravity is important; usually the dense and dilute branches are separated by a large transition parameter space. However, we see in Fig. \ref{fig:MvsR} that this assumption is badly violated as $f$ grows above $\L_{\rm GUT}$. At $\delta=10^{-1}$ (purple curve), the high-$\D$ cutoff occurs very close to $M_{\rm max}$; we were unable to push to larger $\d$ in this analysis due to the appearance of higher-order corrections, but it is plausible that for $\d=\Ocal(1)$ the naive maximum mass is never reached. Other aspects of the solutions in Fig. \ref{fig:MvsR} will be discussed in the next section. 

An important consequence of Eq.~\eqref{eq:GRB} for large $f$ is that due to shifts in the parameter $\D$ as $f$ increases, decay processes can be relevant even on the dilute and transition branches. We address this in the next section.

\section{Decay Rate} \label{sec:decay}

Axion stars decay through emission of relativistic particles. There are two classes of decay modes, which proceed through tree-level \cite{Eby:2015hyx,Mukaida:2016hwd,Eby:2017azn} and loop-level \cite{Braaten:2016dlp} diagrams\footnote{The tree and loop-level processes are sometimes referred to as \emph{classical} and \emph{quantum} decay, respectively.}.
Importantly, the momentum distribution of bound axions has a finite width, which allows for tree-level decay processes that would be forbidden by energy-momentum conservation for particles in momentum eigenstates \cite{Eby:2015hyx,Eby:2017azn}. For a $\phi^n$ potential, the leading process of this type is $(n-1)\,a_c \to a_f$, where $n-1$ bound (``condensed") axions $a_c$ annihilate to a single relativistic axion $a_f$ emitted from the star. We focus below on this tree-level decay process, though of course our conclusions would only be made stronger if higher-order processes were included as well. 

In the specific case of the axion potential in Eq. \eqref{Vphi}, the leading self-interaction potential at $\Ocal(\phi^4)$ gives rise to a $3\,a_c\to a_f$ decay process. One can approximate the decay rate $\G_3$ for this process 
 by taking the matrix element of the self-interaction potential, between the initial state $\langle N\,a_c\vert$ and final state {$\vert (N-3)\,a_c + 1 a_f\rangle$}. The outgoing particle is emitted as a spherical wave, conserving average momentum for the [remaining condensate $+$ free particle] system (see \cite{Eby:2017azn} for a detailed discussion). 
 
 Previous work on relativistic decay processes in axion stars \cite{Eby:2015hyx,Levkov:2016rkk,Visinelli:2017ooc,Eby:2017azn} have assumed the decoupling of gravity in the region where decay becomes relevant.
However, we find in this work that gravity does not decouple in this way when $f$ becomes large, and thus we repeat the calculation of the lifetime in the Appendix \ref{app:rate}, including both effects.

A second assumption made in previous work is that the axion star can track its equilibrium configuration as it decays, which we refer to as the \emph{adiabatic approximation}. However, simulation results \cite{Levkov:2016rkk,Visinelli:2017ooc}, previous semi-analytic estimations \cite{Eby:2015hyx,Eby:2017azn}, as well as the present work (see Appendix \ref{app:rate}), all suggest that when decay becomes relevant, the resulting explosion of relativistic particles is extremely rapid, likely to greatly outpace the relaxation of the star back to equilibrium. Therefore in this work, we argue that an \emph{instantaneous approximation} for the lifetime of axion stars is more appropriate. We thus define the lifetime as
\begin{equation}
 \tau_0 = \frac{N}{3}\frac{1}{\G_3},
\end{equation}
following Eq. (4.1) of \cite{Eby:2015hyx} evaluated in the instantaneous limit\footnote{In previous work \cite{Eby:2015hyx,Eby:2017azn}, we found in the adiabatic approximation that the decay rate scales roughly as $\tau \propto (\D^2/m)\,\exp(c/\D)$ (where $c$ is a constant); in this paper we instead use the instantaneous approximation and find $\tau \propto (\D/m)\,\exp(c/\D)$, reproducing results previously obtained using classical field theory \cite{Hertzberg:2010yz,Grandclement:2011wz}. In both cases the prefactor retains no explicit dependence on $f$. We clarify this point in detail in the Appendix \ref{app:rate}, as it has led to some misinterpretations in the recent literature \cite{Hertzberg:2020xdn}.}.
This lifetime can be uniquely determined for each configuration (on any branch of solutions), which we compare to the age of the universe $\tau_U \approx 13.8\times10^{9}$ years. 

As described in the Appendix \ref{app:rate}, we find that there is a sharp transition from $\tau_0 \gg \tau_U$ (stability) to $\tau_0 \ll \tau_U$ (instability) over a narrow range of $M$, which we define $M_\tau(\d)$. In Fig. \ref{fig:MvsR}, we mark $M_\tau$ by the change from solid (stable, $\tau > \tau_U$) to dashed (unstable, $\tau<\tau_U$) lines on each curve. Importantly, at large enough $f$, the critical point $M_\tau$ occurs on the dilute branch, which has until now been regarded as perfectly stable. 

The precise value of $M_\tau$ depends exponentially on $\D$ but only polynomially on $m$. Owing to this functional dependence, the transition from $\t < \tau_U$ to $\t>\t_U$ occurs at a very sharply defined $M_\tau$, varying only by at most a factor of $2$ as the axion mass $m$ varies between $10^{-20}- 1$ eV.

We further illustrate in Fig. \ref{fig:Mcrit} how the decay instability (blue shaded region) sets in on the dilute branch, at lower mass than the onset of gravitational instability (black shaded region), when $\d\gtrsim 10^{-3}$ ($f\gtrsim 8\times10^{16}$ GeV). The limit of very large binding energy, where $\D =1/\sqrt{2}$, where our approximation is no longer valid, is illustrated by the green shaded region. The blue band represents $M_\tau(\d)$, and its width represents its variation upon varying $m$ in the range $10^{-20}-1$ eV. The red band similarly represents the onset of decay instability on the transition branch of solutions for smaller $\d$.

\section{Discussion}

In this work, we have analyzed the structure and stability of axion stars when the decay constant $f$ is large, with a focus on the approach to the Planck scale $M_P$. Such scenarios are motivated by a significant body of literature, and previous work has typically assumed that results at smaller $f$ can be extrapolated to large $f$. We have shown that this extrapolation is erroneous; at the order of $\D$ that we consider, the dense and transition branches of solutions no longer exist
above $\d \gtrsim 10^{-1}$ ($f \gtrsim 8\times10^{17}$ GeV) up to very large $\D = 1/\sqrt{2}$, and the decay instability point $M_\tau$ occurs on the dilute branch for $\d \gtrsim 10^{-3}$ ($f \gtrsim 8\times10^{16}$ GeV), as shown in Fig. \ref{fig:MvsR}. 

This result has applications in any scenario where axion stars form with $f\gtrsim10^{17}$ GeV, including those outlined in the introduction. For example, for ULDM with particle mass $m\sim 10^{-22}$ eV, the correct relic abundance is obtained for $f\sim10^{17}$ GeV \cite{Hui:2016ltb}. 
ULDM simulations generally find axion stars forming the cores of galaxies in this scenario \cite{Schive:2014dra,Schive:2014hza,Mocz:2017wlg,Veltmaat:2019hou,Nori:2020jzx}, with masses that are safely below the instability points we find here; however, if the axion star masses had been a factor of $\sim 80$ larger at formation, or if they can accrete mass efficiently, then ULDM axion stars would not merely collapse but also decay on the dilute branch, strengthening the argument of previous studies \cite{Eby:2018zlv} (see Appendix \ref{app:comp} for details). Such an effect would not be seen in standard ULDM simulations, as they typically neglect both the self-interaction potential as well as relativistic effects. As simulations of axion star formation and accretion become more precise, or as nonminimal axion models are investigated, this must be taken into account in the final analysis.

Further, phenomenological studies of axion stars (including those looking for gravitational wave signals \cite{Clough:2018exo,Dietrich:2018jov} or new black hole formation mechanisms \cite{Helfer:2016ljl,Chavanis:2017loo,Michel:2018nzt,Widdicombe:2018oeo}) must acknowledge that decay may make their proposed configurations unstable. For $f \simeq 7\times10^{17}$ GeV, the mass of truly stable axion stars is already a factor of $\sim 3$ lower than the usual boundary of gravitational stability, as shown in Fig. \ref{fig:Mcrit}. This in fact precludes axion stars with neutron star-like masses and radii by nearly an order of magnitude.

We should point out, of course, that our own results should not be naively extrapolated either. For example, in the limit $f\to\infty$ the axion self-interactions will decouple, and in that limit neither nonrelativistic collapse nor relativistic decay processes discussed here destabilize the star. Taking $f>M_P$ may, however, be in tension with theoretical considerations like the Weak Gravity Conjecture (see e.g. \cite{Montero:2015ofa}), though there may be ways to reconcile them \cite{Kaplan:2015fuy,Fonseca:2019aux}.

Previous works have also considered decay modes other than $3\to1$, including so-called \emph{quantum} decay (e.g. $4\to2$) \cite{Hertzberg:2010yz}, other classical decay processes (e.g. $5\to1$) for alternate axion potentials \cite{Zhang:2020bec}, or for repulsive self-interactions \cite{Hertzberg:2020xdn}. In some contexts it is claimed that the resulting decay modes can actually dominate the decay rate. Because we have not included such contributions in this work, we emphasize that the lifetime we calculate is merely an upper bound on the true axion star lifetime.

We found in our study that higher-order corrections to the EKG equations appear not only at higher orders in $\D^2$ (as found in \cite{Eby:2017teq}), but also as powers of the product $\d\,\D^2$ and as $\d^2$; as a result we were not able to robustly analyze configurations with $\d \gtrsim 0.1$, corresponding to $f \gtrsim 8\times10^{17}$ GeV. We leave the effect of these corrections, and the (in)stability of axion stars with $\d = \Ocal(1)$, for a future study, It would also be worthwhile to make a thorough comparison of our results, which apply to static axionic configurations, to the dynamical works of \cite{Helfer:2016ljl,Widdicombe:2018oeo} which use very different initial conditions; see Appendix \ref{app:comp} for a more heuristic comparison.

\section*{Note Added}

After completion of this project, a related work \cite{Zhang:2020ntm} appeared, which illustrates a different method but comes to many of the same conclusions. We believe the two papers complement one another.

\section*{Acknowledgements}

We are grateful to M. Amin, M. Hertzberg, M. Leembruggen, and H.-Y. Zhang for helpful contributions and comments on this work.
The work of J.E. was partially supported by the Zuckerman STEM Leadership Fellowship and by World Premier International Research Center Initiative (WPI), MEXT, Japan.
L.S. and L.C.R.W. thank the University of Cincinnati Office of Research Faculty Bridge Program for funding through the Faculty Bridge Grant. 
L.S. also thanks the Department of Physics at the University of Cincinnati for financial support in the form of the Violet M. Diller Fellowship.



\appendix


\section{Detailed Calculation of the Decay Rate}  \label{app:rate}

\subsection{Position of the Singularity at $\k\neq 0$}

To analyze decay processes, in previous work some of us showed how to extend the RB field operator to include a scattering state contribution in Eq. (\ref{RB}), allowing axion quanta with energy $\o_p>m$ \cite{Eby:2015hyx,Eby:2017azn}. The scattering state wavefunction $\psi_f$, at leading order in spherical harmonics, is given by
\begin{equation} 
 \psi_f(t,r) = \frac{1}{2\pi^2} \int_0^\infty \frac{dp\,p}{2\o_p}\,j_0(p\,r)\,e^{-i\,\o_p\,t}\,a_{00}(p),
\end{equation}
where $a_{00}(p)$ 
is the annihilation operator for a scattering state of momentum $p$ labeled by its angular momentum quantum numbers $\ell=\ell_z=0$, and $j_0$ is the zeroth spherical Bessel function.
The decay rate was then analyzed for an attractive $\phi^4$ potential, where the rate is always nonzero due to an essential singularity in the equation of motion; however, the matrix element is exponentially suppressed for weakly-bound axion stars. In the range of parameters considered there (for example, $f\sim 10^{10-12}$ GeV), the decay process was irrelevant on the dilute branch and became important on the transition branch. The lifetime of an axion star on the transition branch was found to be
\begin{equation} \label{eq:lifetime} 
 \tau = \frac{3\,y_t}{4096\pi^3\,y_I^3}\frac{\D^2}{m}e^{4\sqrt{2}y_I/\D},
\end{equation}
where the constant $y_t$ was determined by fitting the curve $M(\D)$ on the transition branch, and the constant $y_I$ characterized the position of the singularity in the complex plane at $x = i\,y_I$

\begin{figure*}[t]
\centering
\includegraphics[scale=0.42]{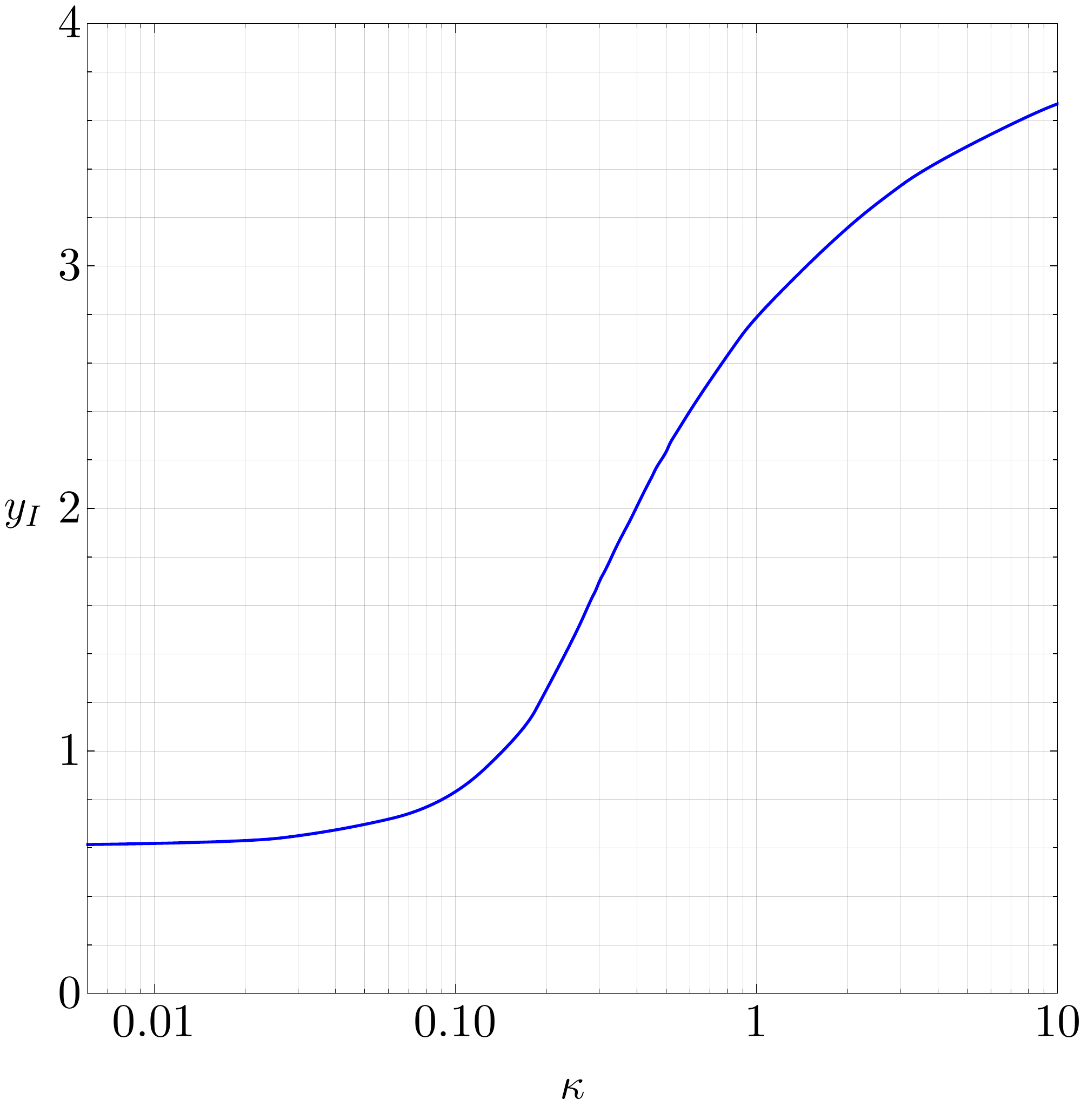}
\includegraphics[scale=0.58]{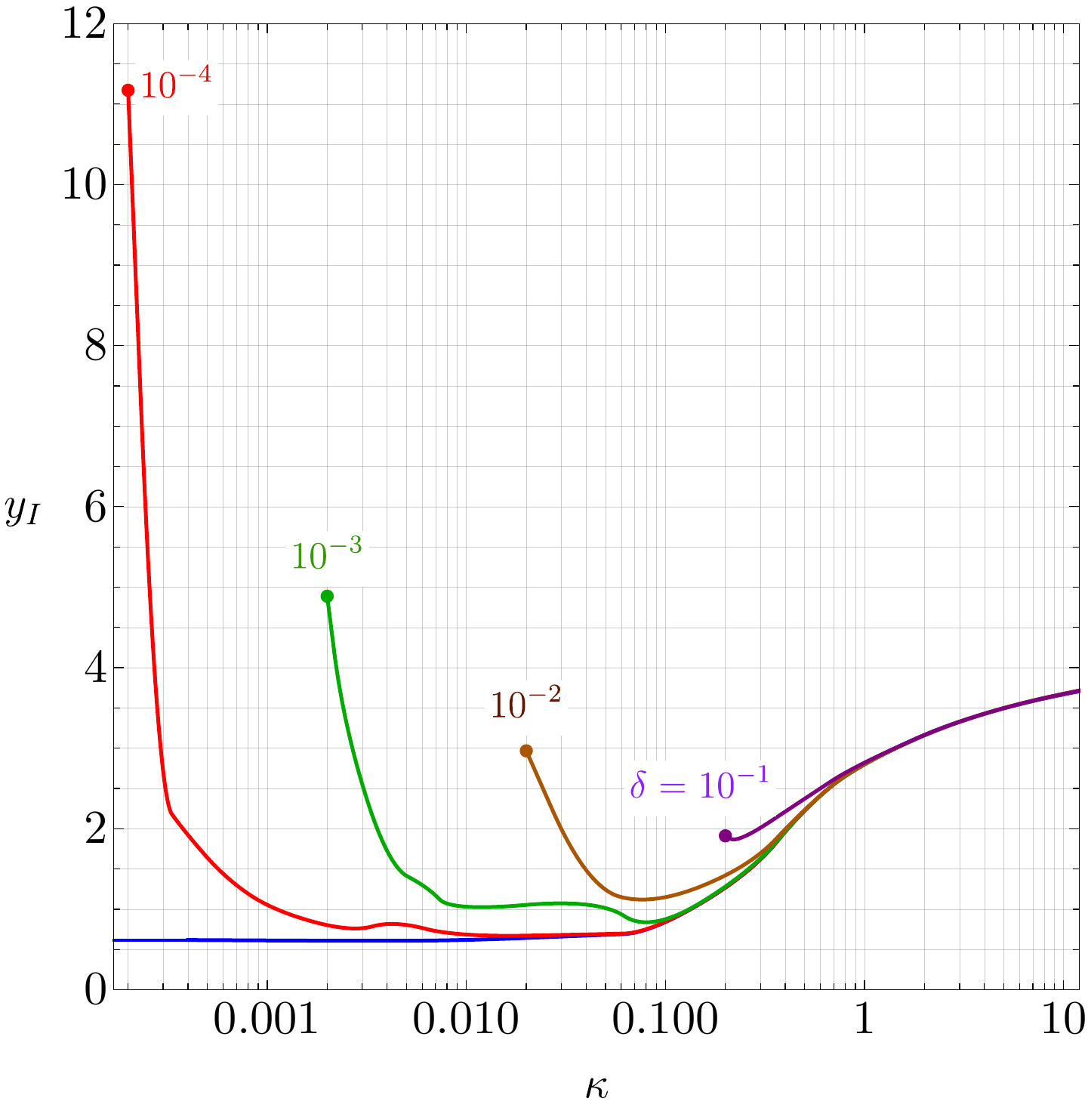}
\label{fig:singularity}
\caption{Position of the singularity $y_I$ as a function of the effective gravitational coupling $\kappa$. \\
\underline{Left:} The position of the singularity at leading-order in RB, which converges to a value $y_I = 0.602$ when $\kappa \ll 1$. \\
\underline{Right:} The position of the singularity at leading-order in GRB, for large $f$ approaching $M_P$; the blue, red, green, brown, and purple correspond to log$_{10}(\d) = -6,-4,-3,-2,-1$ respectively. For sufficiently small $f$, $y_I$ converges to the value $y_I = 0.602$ at small $\k$. The filled dots correspond to a cutoff of the solutions at the value $\D = 1/\sqrt{2}$.}
\end{figure*}

The lifetime in Eq. \eqref{eq:lifetime} is approximately correct in its region of applicability, but rests on assumptions about the input parameters. For example, in \cite{Eby:2015hyx,Eby:2017azn} it was assumed that the effect of gravity could be neglected (using the $\k\to0$ limit of Eq. (\ref{EKGY})), which becomes appropriate near $M\approx M_{\rm max}$ and remains so on the transition branch. Then the Klein-Gordon equation for $Y(x)$ has a unique solution corresponding to $Y(0) = 12.268$, and the constant value for $y_I \approx 0.602$ is uniquely determined. Because gravity does not decouple in this way at large $f$ (see Main Text), we determine the position of the singularity for nonzero $\k$ below. We also show how the singularity is shifted to larger values of $y_I$ at leading-order in GRB, leading to slower decay rates than one would have obtained using the constant $y_I \approx 0.6$. The calculation is detailed below, and the result is given in Fig. \ref{fig:singularity}.

The decay rate depends upon the integral \cite{Eby:2015hyx}
\begin{align} \label{I3full} 
 I_3(\D) = \frac{1}{\D^2}\int_{-\infty}^{\infty} dx\,x\,\exp\left(\frac{i\,k_3\,x}{\Delta}\right)\,J_3\left[\D\,Y(x)\right] \nn \\
     \qquad \qquad  {\rm (``Bessel")},
\end{align}
which, at small enough $\D$, can be approximated by
\begin{equation} \label{I3} 
 I_3(\D) = \frac{\Delta}{48}\int_{-\infty}^{\infty} dx\,x\,\exp\left(\frac{i\,k_3\,x}{\Delta}\right)\,Y(x)^3
  \qquad     ({\rm ``}\,Y^3\,{\rm"}),
\end{equation}
where $k_3=\sqrt{9\e^2-1}\simeq \sqrt{8}$ is the momentum of the outgoing relativistic axion for a $3\,a_c\to a_f$ annihilation, in units of the axion mass. 
These integrals, labeled for future convenience, can be calculated directly by numerical integration only at relatively large $\D$, as the integrand is highly oscillatory at small $\D$. 

A simpler approach is available, as the integration is dominated by the leading singularity of the wavefunction $Y(x)$ in the complex plane. To determine the position of this singularity, we follow the prescription of \cite{Eby:2015hyx} and use the following ansatz for the wavefunction in the vicinity of the singularity:
\begin{equation} \label{ansatz} 
 Y_s(x) = \frac{A}{x^2 + y_I{}^2}.
\end{equation}
When $\k>0$, the gravitational potential $b(x)$ shifts the position of the singularity, and must be determined self-consistently with the wavefunction. 

To determine the effect of gravity on the singularity, observe that under spherical symmetry, we can rewrite Eq.~\eqref{EKGb} as
\begin{align} 
 b(x) &= -\frac{1}{8\pi}\left[\int d^3 x'  \frac{Y(x')^2}{|\vec{x}-\vec{x}'|}\right] \nn \\
	&= -\frac{1}{8\pi}\int dx'\,x'^2\,Y(x')^2\int dc_\theta d\phi \frac{1}{\sqrt{x^2+x'^2 - 2\,x\,x'\,c_\theta}} \nn \\
	&= -\frac{1}{4}\left[\int_0^x dx'\frac{x'^2\,Y(x')^2}{2\,x} + \int_x^\infty dx'\,\frac{x'\,Y(x')^2}{2}\right].
\end{align}
Then substituting the ansatz of Eq.~\eqref{ansatz} for $Y(x)$ we can perform the integrals explicitly; we find 
\begin{align} 
 b_s(x) &= -\frac{A^2}{16}\frac{\tan^{-1}\left(\frac{x}{y_I}\right)}{x\,y_I}.
\end{align}
Of course, the combination $Y_s$ and $b_s$ satisfies the Poisson equation \cite{Eby:2018dat}
\begin{equation} 
 \nabla_x^2\,b_s(x) = \frac{1}{2}Y_s(x)^2.
\end{equation}
Thus we can see that the solution for $b(x)$ near the singularity at $x=i\,y_I$ is regular.

We now Taylor expand the wavefunction and gravitational potential around $x=0$, as
\begin{equation}  \label{Ytaylor} 
 Y(x) = \sum_{n=0}^\infty \eta_n\,x^{2n}, \qquad
 b(x) = \sum_{n=0}^\infty \beta_n\,x^{2n} .
\end{equation}
Evaluating the equations of motion (\ref{EKGY}-\ref{EKGb}), we obtain recursion relations among the expansion parameters $\eta_n$ and $\beta_n$; for example, the coefficient multiplying $x^{2n}$ in Eq.~\eqref{EKGY} must evaluate to zero, which implies
\begin{align}
(2n+2)(2n+1)\eta_{n+1} &+ 4(n+1)\eta_{n+1} \nn \\
				&+ \frac{1}{8}\sum_{i,j\geq0}\eta_i\eta_j\eta_{n-i-j} - \eta_n \nn \\
 			&- \kappa\sum_{i\geq0}\beta_i\,\eta_{n-i} = 0,
\end{align}
and similarly for $\beta_n$ and Eq.~\eqref{EKGb}. We therefore leave $\eta_0$ and $\beta_0$ as free parameters and iteratively solve for $\eta_{n>0}$ and $\beta_{n>0}$ using these recursion relations, truncating at some large maximum integer in the expansion. The first-order terms in the expansions imply the coefficients
\begin{align} \label{eq:eta1}
 \eta_1 &= \frac{1}{6}\left[(1 + \beta_0\,\kappa)\eta_0 - \frac{\eta_0^3}{8}\right], \nn \\
 \beta_1 &= \frac{\eta_0^2}{12}, \nn
\end{align}
whereas the second order implies
\begin{align}
 \eta_2 &=\frac{1}{7680}\left[64\eta_0(1 + \beta_0\,\kappa)^2 - 32\eta_0^3(1 - \kappa + \beta_0\,\kappa) + 3\eta_0^5\right], \nn \\
 \beta_2 &= \frac{1}{960}\left[8\eta_0^2(1 + \beta_0\,\kappa) - \eta_0^4\right], \nn
\end{align}
and so on.

We can then match Eq.~\eqref{ansatz} and Eq.~\eqref{Ytaylor} to obtain relations for $A$ and $y_I$ in terms of the parameters $\eta_n$; at $n$th order in the expansion, we obtain \cite{Eby:2015hyx}
\begin{align}
 A &= (-1)^n\,\eta_n\,y_I^{2n+2}, \\
 y_I &= \sqrt{-\frac{\eta_{n-1}}{\eta_{n}}}.
\end{align}
Noting that using the recursion relations above, we have $\eta_n=\eta_n(\eta_0,\beta_0,\kappa)$, we can take a given solution and evaluate the position of the singularity $y_I(\eta_0,\beta_0,\kappa)$. In the limit $\kappa\to0$ (gravitational decoupling), we recover the result of \cite{Eby:2015hyx} that $A=8\,y_I$ and $y_I = 0.602$. More generally at non-zero $\k$, we find $A=8\,y_I$ still holds, and illustrate the position of the singularity $y_I$ in the left panel of Figure \ref{fig:singularity}.

\subsection{Position of the Singularity in GRB}

We turn now to the structure of the essential singularity including relativistic corrections, using the Generalized Ruffini-Bonazzola (GRB) formalism. The equation of motion for $Y(x)$ at next-to-leading order in the GRB formalism is Eq.~\eqref{eq:GRB}.
Using the recursion procedure above, we can determine the singularity structure of this equation as well. There are small modifications to the coefficient relations, e.g. $\eta_1$ in Eq.~\eqref{eq:eta1} is modified at $\Ocal(\D^2)$ as
\begin{equation}
 \eta_1 = \frac{1}{6}\left[(1 + \beta_0\,\kappa)\eta_0 - \frac{\eta_0^3}{8} + \frac{3}{512}\D^2\,\eta_0^5\right],
\end{equation}
but the basic procedure is the same as above.

The GRB equation of motion depends on an additional parameter $\Delta$ in addition to $\k$, though at fixed $\D$ we trade the latter for $\d = \D^2\,\k$.
Therefore we characterize our solutions by the input parameters $\{\eta_0,\beta_0,\D,\d\}$.
We solved the GRB equation over a large range inside the bounds $10^{-4} \leq \D \leq 1/\sqrt{2}$ and $10^{-6} \leq \d \leq 10^{-1}$ (this corresponds to $10^{-4} \lesssim \k \lesssim 10^3$); at the largest values of $\D$ we expect the higher-order contributions in GRB to be very relevant, and worse yet, such solutions are unphysical due to extremely high binding energies \cite{Eby:2019ntd}. Still, we can analyze the structure of solutions as an academic exercise, and we will see that solutions with very large values of $\D$ are not phenomenologically relevant anyway due to fast decay rates.

In the right panel of Figure \ref{fig:singularity}, we illustrate the shift in the singularity position for different choices of $\d$, with the goal of approaching $f=\Ocal(M_P)$. We see that, as expected, when $f$ is sufficiently small (given by the blue curve, or $f\lesssim 2.4\times10^{15}$ GeV), the singularity is unaffected by the GRB correction over the full range of $\kappa$ we analyzed. As $f$ increases, the deviation in the singularity point appears at larger values of $\kappa$, though at the same time the cutoff at $\D=1/\sqrt{2}$ (given by the circles in the Figure) reduces the relevant physical range.

\begin{figure*}[t]
\centering
\includegraphics[scale=0.62]{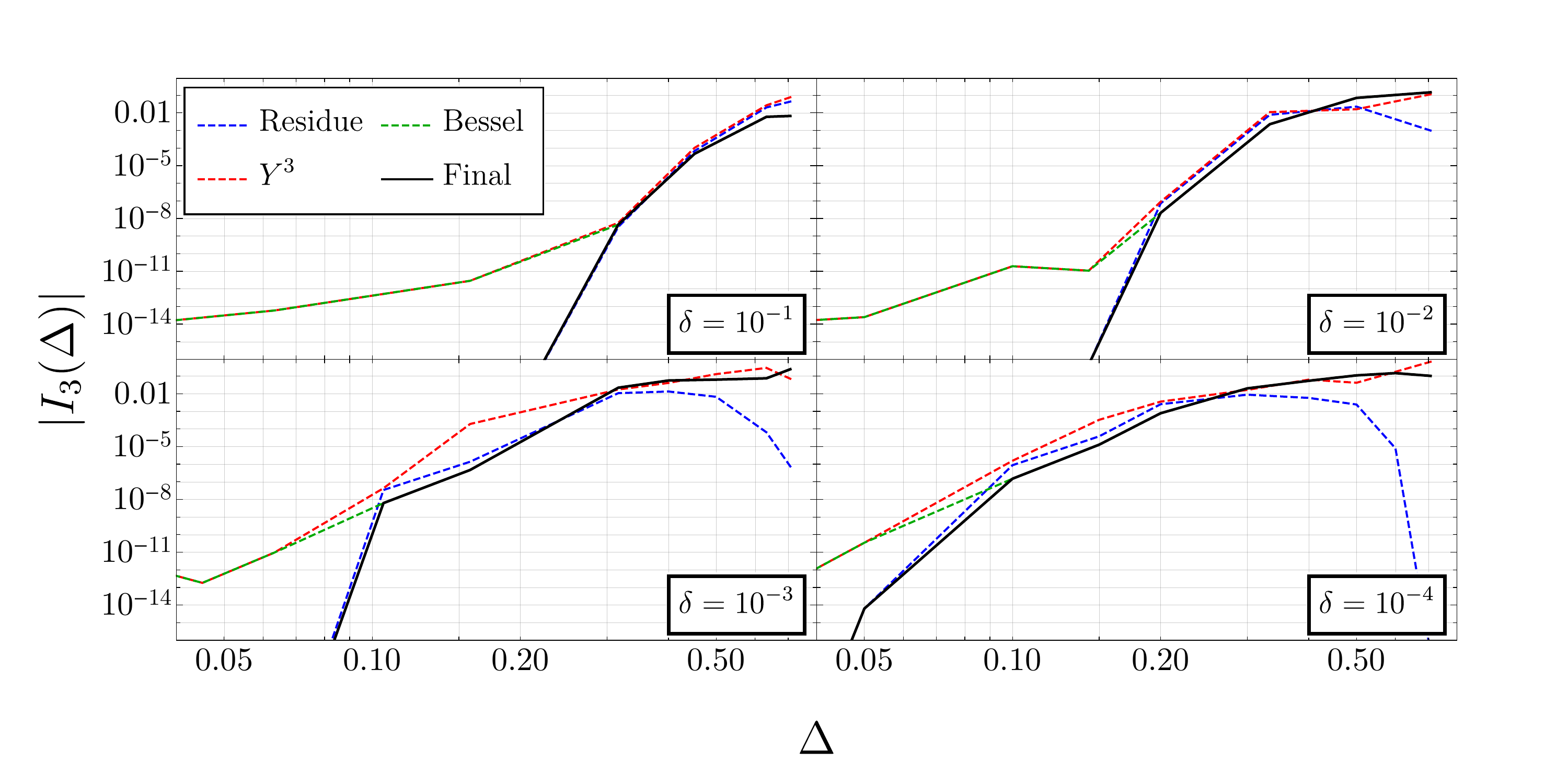}
\label{fig:I3comp}
\caption{Comparison of different calculation methods for $I_3(\D)$: From residue theorem (blue); from numerical integration over $Y^3$ (red); from numerical integration over $J_3(\D\,Y)$ (green); and our final estimation (black) which interpolates between the blue curve at small $\D$ and the green curve at large $\D$. The choices of $\d$ are $\d=10^{-1}$ (top left), $\d=10^{-2}$ (top right), $\d=10^{-3}$ (bottom left), and $\d=10^{-4}$ (bottom right).}
\end{figure*}

We can now proceed to evaluate the lifetime of axion stars at large $f$ and moderate $\D$, at leading-order in GRB. With the position of the singularity $y_I$ in hand, we can use the residue theorem to compute $I_3$ in Eq.~\eqref{I3} and find
\begin{equation} \label{eq:I3_res}
 I_3(\D) \simeq i\,\frac{32\,\pi}{3}\frac{y_I}{\Delta}\exp\left(-\frac{k_3\,y_I}{\Delta}\right)
 \qquad   {\rm (``Residue")},
\end{equation}
which depends exponentially both on the inverse of the binding energy parameter $1/\D$ and the position of the singularity $y_I$. 

However, note that at large $\D$ the result in Eq~\eqref{eq:I3_res} will not be correct, as the application of the residue theorem assumed the validity of Eq.~\eqref{I3}. When the product $\D\,Y(x)$ is large, we must not expand the Bessel function $J_3(\D\,Y(x))$ in Eq.~\eqref{I3full}; fortunately, it is in this range that the integrand does not oscillate fast and we can integrate $I_3$ directly. To briefly summarize:
\begin{itemize}
 \item The analytic ``Residue" result of Eq.~\eqref{eq:I3_res} is applicable at low $\D \lesssim 0.1$;
 \item The numerical result ``Bessel" from integrating $J_3$ in Eq.~\eqref{I3full} is applicable at the largest $\D \sim \D_{\rm max} = 1/\sqrt{2}$;
 \item Both should roughly agree, with each other and with Eq.~\eqref{I3} (``$Y^3$"), in an intermediate range $0.1 \lesssim \D \lesssim \D_{\rm max}$.
\end{itemize}
We illustrate the results of these estimations (blue for residues, green for Bessel, red for $Y^3$) in Figure \ref{fig:I3comp}. The black line, which we use to calculate the decay rate in the next section, interpolates between Eq.~\eqref{eq:I3_res} at low $\D$ and Eq.~\eqref{I3full} at high $\D$; when in doubt we used the smaller estimation of $|I_3|$ to make a conservative estimation of the decay rate.

\begin{figure*}[t]
\centering
\includegraphics[scale=0.6]{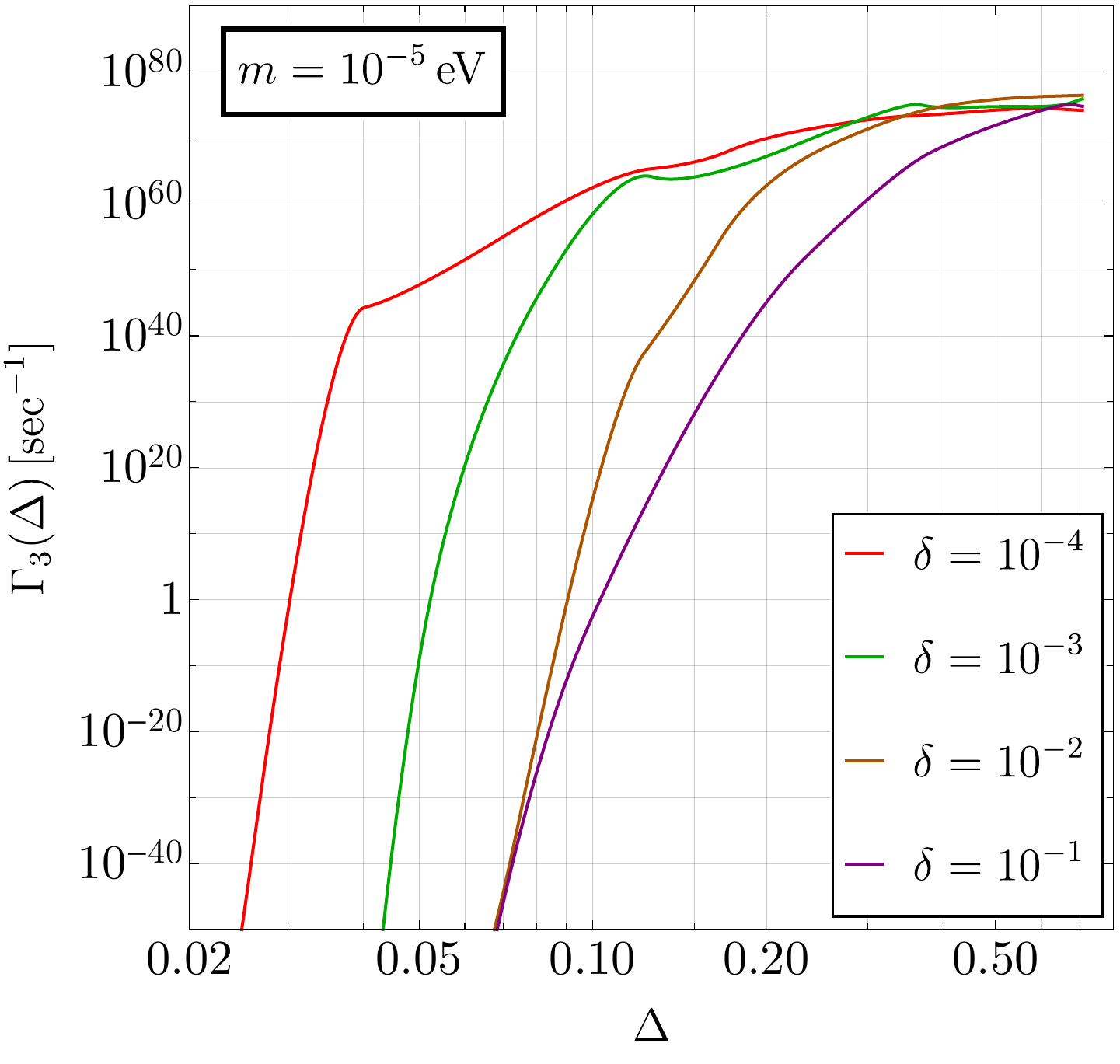}\,\,
\includegraphics[scale=0.6]{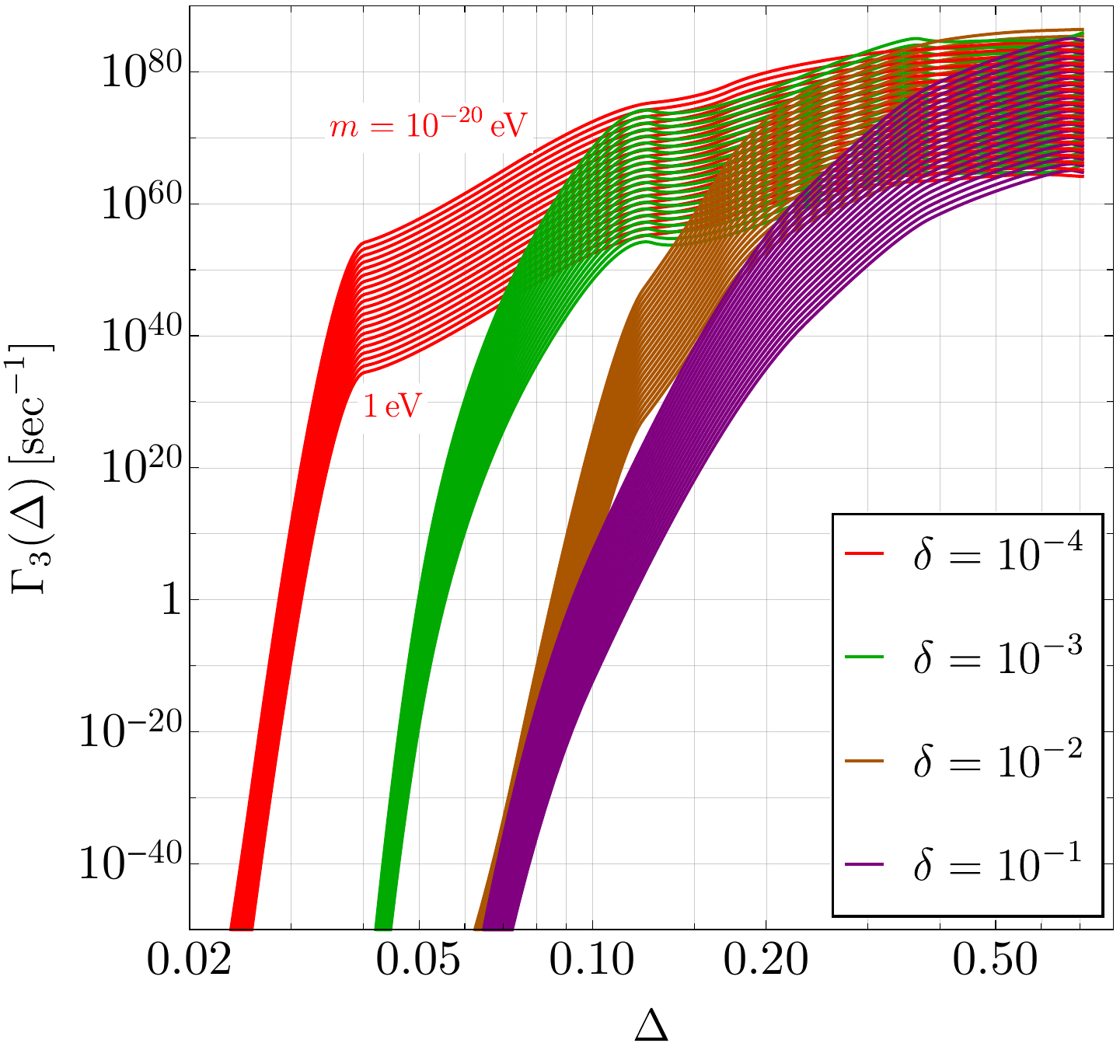}
\caption{The single-annihilation decay rate $\G_3$ for $\d=10^{-4},10^{-3},10^{-2},10^{-1}$ (respectively the red, green, brown, and purple curves). In the left panel, the particle mass is fixed to $10^{-5}$ eV, whereas in the right panel $m$ is varied on each order of magnitude between $10^{-20}-1$ eV (lines from top to bottom of each band).}
\label{fig:G3}
\end{figure*}

\subsection{Decay Rate and Lifetime}

The decay rate for a single annihilation process is given by \cite{Eby:2015hyx,Eby:2017azn}
\begin{equation}
 \G_3(\D) = \frac{f^2}{2\pi k_3\,m}|I_3(\D)|^2.
\end{equation}
We show the resulting rate of annihilations/sec in Fig. \ref{fig:G3} for different choices of $\d=10^{-4},10^{-3},10^{-2},10^{-1}$ (the red, green, brown, and purple curves, respectively). In the left panel we have fixed the axion mass $m=10^{-5}$ eV, whereas we vary $m$ in the right panel between $m=1$ eV (bottom of each band) and $m=10^{-20}$ eV (top). We observe a very rapid increase of the decay rate, from extremely small values $\ll 1$ annihilation/sec to greater than $10^{40}$ annihilations/sec around (for example) $\D\simeq 0.03$ for $\d=10^{-4}$. This increase occurs at slightly higher values of $\D$ for larger $m$ or for larger $f$ (i.e. larger $\d$).

The lifetime is proportional to $|I_3|^{-2} \propto \exp(y_I/\D)$, implying that dilute axion stars (corresponding to the smallest allowed $\D$) may be stable and survive longer than the age of the universe. However, this has only been explicitly investigated at small values of $f$, and we have seen that naive extrapolation of such results to large $f$ may not be appropriate. Indeed, we find that contrary to conventional wisdom about axion stars, the lifetime for $f \gtrsim 10^{17}$ GeV is shorter than the age of the universe \emph{even on the dilute branch}, as explained below. 

\begin{figure}[t]
\centering
\includegraphics[scale=0.6]{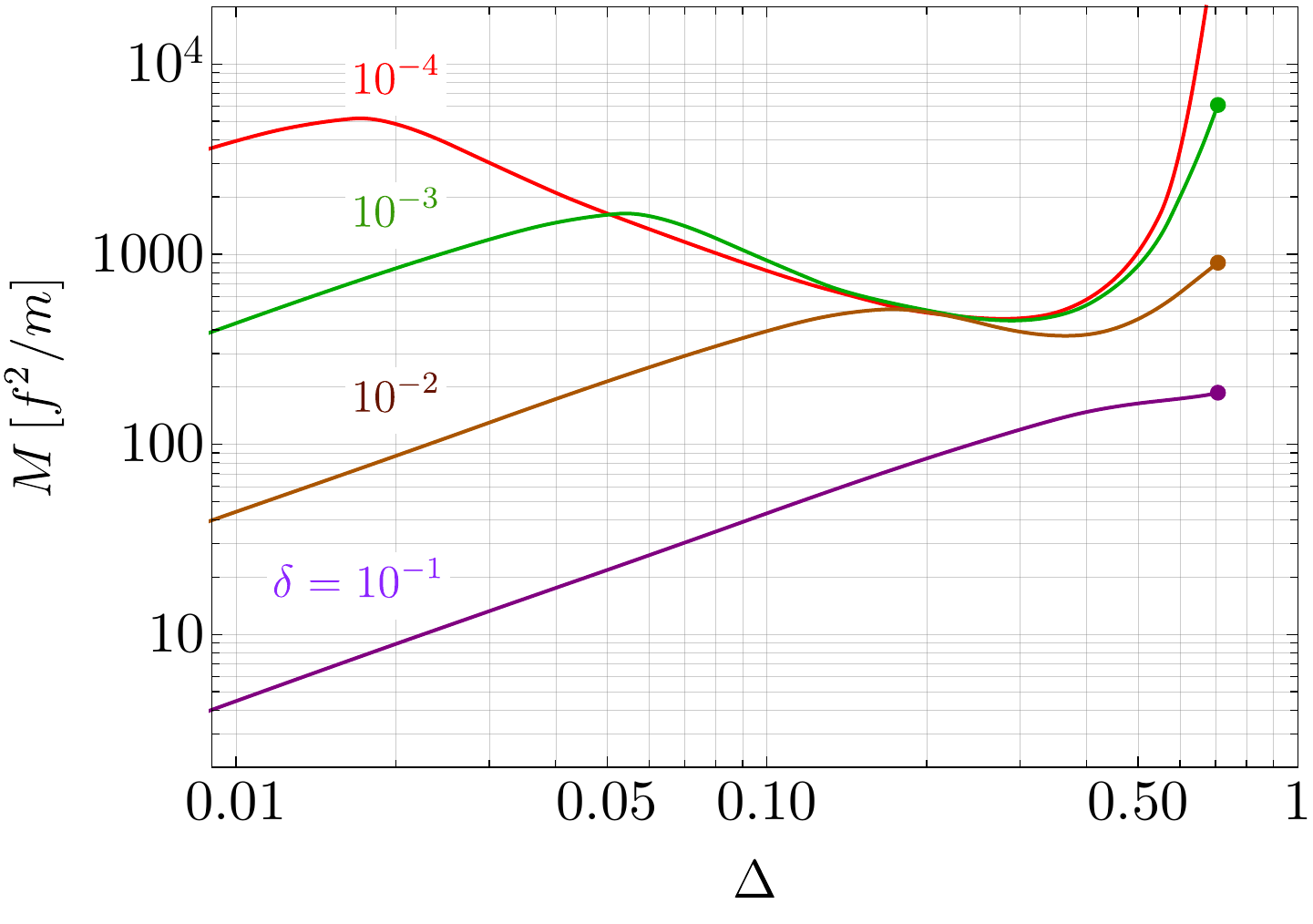}\,
\caption{Mass of axion stars as a function of $\D$, for different choices of $\d$ (colors match those of Figure \ref{fig:G3}). The filled circles represent the endpoint of our set of solutions, at $\D = 1/\sqrt{2}$.}
\label{fig:MvsDelta}
\end{figure}

We estimate the lifetime numerically as follows. In a region of parameter space where the decay rate is large, we may approximate it as constant, as the axion star explodes in a rapid Bosenova of relativistic particles; we call this the \emph{instantaneous approximation}. In that case (assuming a constant decay rate), the timescale for the evaporation of the star would be
\begin{equation} \label{tau0}
 \tau_{0} = \frac{N}{3}\,\frac{1}{\G_3} \simeq \frac{2\pi\,k_3}{3\,f^2}\frac{M(\D)}{|I_3(\D)|^2},
\end{equation}
where the prefactor $N/3$ is the number of annihilations necessary to completely deplete the star (for each annihilation, $3$ bound axions are lost).
Then, because $M$ has a one-to-one relationship with $\D$, we can compute the decay timescale uniquely as a function of $\D$. In practice we use the full solution for $M(\D)$ in the numerical results, which is illustrated in Fig. \ref{fig:MvsDelta}. 
Note that the lifetime in Eq.~\eqref{tau0} is identical to the one used in \cite{Eby:2015hyx,Eby:2017azn}, but here it is evaluated in the instantaneous limit; we clarify the difference at the end of this section.

Using fits on each branch of axion stars, we obtain an analytic form for the lifetime which may help guide the reader's intuition. Firstly, in \cite{Eby:2015hyx,Eby:2017azn}, we noted that (at small $f \simeq 10^{12}$ GeV) the decay rate was exponentially small on the dilute branch and that the lifetime only became short on the transition branch; in that case, the relationship between $M$ and $\D$ is given by
\begin{equation}
 M(\D) = \frac{y_t}{\D}\frac{f^2}{m}
      \qquad \qquad  {\rm (Transition\,\,branch)},
\end{equation}
with $y_{t} \simeq 75.4$ is determined by fitting the mass function shown in Fig. \ref{fig:MvsDelta} \cite{Eby:2017azn}.
Then the timescale for instantaneous decay is
\begin{align} \label{eq:tauINtrans}
 \tau_0 \simeq \frac{2\pi\,k_3\,y_t}{3\,m\,\D}\frac{1}{|I_3(\D)|^2} &\approx \frac{447}{m\,\D}\frac{1}{|I_3(\D)|^2} \nn \\
      &{\rm (Transition\,\,branch)}.
\end{align}

\begin{figure*}[t]
\centering
\includegraphics[scale=0.90]{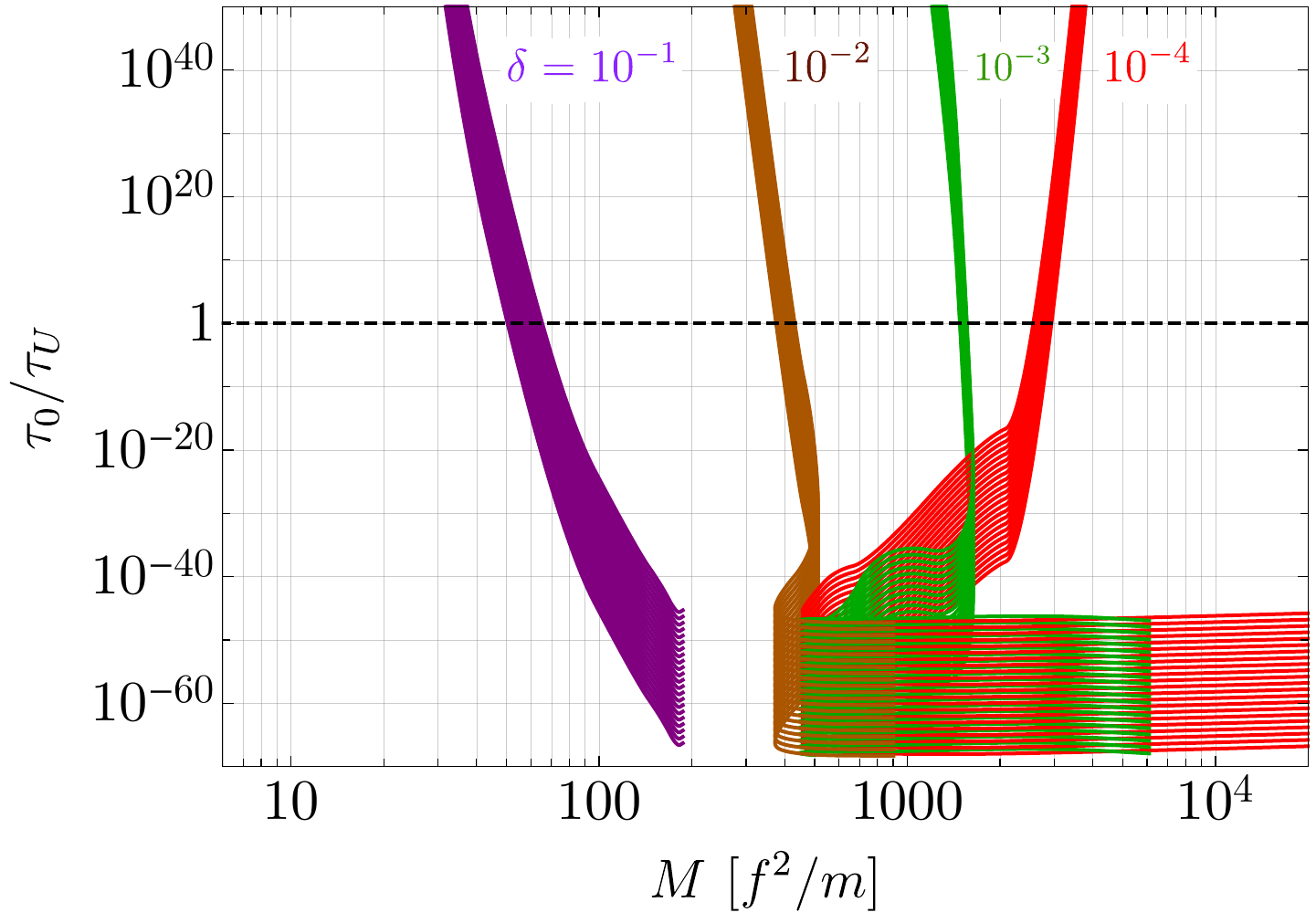}\,
\caption{Lifetime of axion stars (in the instantaneous approximation), normalized to the universe lifetime $\tau_U$, as a function of total mass $M$, for different choices of $\d$ (colors match those of Figure \ref{fig:G3}). The width of the curves represents choices of $m$ varied on each order of magnitude in the range $10^{-20}-1$ eV (lines from bottom to top of each band).}
\label{fig:lifevsM}
\end{figure*}

For large $f$, we find in this work that the decay timescale can, on the dilute branch, already be lower than the age of the universe, as $\D$ becomes large before the crossover point. In that case the mass function is
\begin{equation}
 M(\D) = y_d\,\D\,\frac{f^2}{m}
      \qquad \qquad  {\rm (Dilute\,\,branch)}
\end{equation}
with $y_d = 1.75\times8\pi/\d$ (found by fitting the curves in Fig. \ref{fig:MvsDelta}),
and the lifetime takes the form
\begin{align}
 \tau_0 \simeq \frac{2\pi\,k_3\,y_d\,\D}{3\,m}\frac{1}{|I_3(\D)|^2} \approx \frac{261\,\D}{m\,\d}\frac{1}{|I_3(\D)|^2} \nn \\
      \qquad \qquad  {\rm (Dilute\,\,branch)}.
\end{align}

Due to the rapid turn-on of the decay rate, aside from a very narrow region near $\tau_0 \simeq \tau_U$, the lifetime of axion stars is either (i) so long that the star can be treated as stable with a conserved particle number $N$, or (ii) so short that the star decays almost instantly. In the latter case, the instantaneous approximation above seems very well-justified, as the relaxation time for the axion star to remain in its equilibrium configuration at each instant as it evolves will likely be much longer than the decay timescale $\tau_0$. Therefore, we use the instantaneous case to define the crossover from stability to instability under decay which we describe in the Main Text. The result for the instantaneous lifetime as a function of $M$ and $\d$ is given in Fig. \ref{fig:lifevsM}

\section{Comparison to Previous Work} \label{app:comp}

Below, we review previous work on the decay of axion stars, as well as their possible collapse to black holes, and compare these results with our own.

{\bf \underline{Decay Rate:}} In this work we computed the lifetime of axion stars using an instantaneous approximation. Although we feel this approximation is justified, the arguments above fall short of a proof, as the relaxation timescale has not been worked out in detail. In a situation in which the axion star tracks its equilibrium configuration as it decays, one should integrate the decay rate as a function of $M$ from an initial $M_0$ to some smaller final value $M_f$, i.e.
\begin{equation} \label{eq:tauAD}
 \tau = \frac{1}{3}\int\frac{dN}{\G_3} \simeq \frac{2\pi\,k_3}{3\,f^2}\int_{M_{0}}^{M_f}\frac{dM}{|I_3(\D)|^2}.
\end{equation}
This formulation was used by us in previous work to determine the lifetime of axion stars on the transition branch \cite{Eby:2015hyx,Eby:2017azn}. We found that on the transition branch, $M \propto 1/\D$ but $I_3(\D) \propto \exp\left(-1/\D\right)$, and so $dM/|I_3(\D)|^2 \ggg1$ when $\D \ll 1$. In that case, the lifetime integral is dominated by the smallest $M$ in the integration range (that is, $M_f$), and the result was given by Eq.~\eqref{eq:lifetime} with $M=M_f$. 

An important difference between the adiabatic and instantaneous approximations is the scaling of the lifetime with the parameter $\D$. For example, on the transition branch (where gravity decouples) at small $\D$ (where Eq. \eqref{eq:I3_res} is appropriate), the adiabatic case in Eq. \eqref{eq:tauAD} gives $\tau \sim a_A\,(\D^2/m)\,\exp(c/\D)$, where $a_A$ and $c$ are constants, as found in \cite{Eby:2015hyx,Eby:2017azn} (see Eq. \eqref{eq:lifetime}). On the other hand, the instantaneous case of Eq. \eqref{eq:tauINtrans} gives $\tau \sim a_I\,(\D/m)\,\exp(c/\D)$, where $a_I$ is another constant. The latter is precisely the scaling found in previous investigations using classical field theory; see e.g. Eq. (24) of \cite{Hertzberg:2010yz} or Eq. (50) of \cite{Grandclement:2011wz}. In all cases, the prefactor retains no explicit dependence on $f$.

{\bf \underline{ULDM Simulations:}} In these simulations \cite{Schive:2014dra,Schive:2014hza,Mocz:2017wlg,Veltmaat:2019hou,Nori:2020jzx}, axion stars form with very large masses $M$ due to the smallness of the particle mass $m\simeq 10^{-22}$ eV; in large halos of total mass $M_h \gtrsim 10^{9} M_\odot$, the axion star mass appears to follow a now well-known relation \cite{Schive:2014dra,Bar:2018acw}
\begin{equation}
 M = 1.4\times10^{9}M_\odot\left(\frac{10^{-22}\,{\rm eV}}{m}\right)\left(\frac{M_h}{10^{12}M_\odot}\right)^{1/3}.
\end{equation}
Comparing to the maximum gravitationally-stable mass $M_{\rm max} = 10.2\,M_P\,f/m$, we see that
\begin{equation}
 \frac{M}{M_{\rm max}} \simeq 0.013\left(\frac{10^{17}\,{\rm GeV}}{f}\right)\left(\frac{M_h}{10^{12}M_\odot}\right)^{1/3},
\end{equation}
so in the largest halos simulated, where $M_h \simeq 10^{12}M_\odot$, a typical ULDM candidate with $f=10^{17}$ GeV will form an axion star core with mass $\sim80$ times below its maximum mass. If axion stars can accrete enough mass after formation to cover this gap, they run the risk of not only collapsing but also decaying, because as illustrated in Fig. \ref{fig:MvsR} the decay instability sets in also in the same parameter range.

{\bf \underline{Collapse Simulations:}} Other simulations have probed the fate of axion stars at large $f = \Ocal(M_P)$ by evolving the classical equations of motion \cite{Helfer:2016ljl,Widdicombe:2018oeo}. One of the purposes of these simulations was to determine under which conditions an axion star might collapse directly to a black hole; their analysis suggests a triple point in the axion star phase diagram at $f_{TP} \simeq 0.3\tilde{M}_P$ and $M_{TP} \simeq 2.4\tilde{M}_P^2/m$, which is $\d_{TP} = 0.09$ and $M_{TP}  = (2.4/\d)(f^2/m) = 26.6\,f^2/m$ in the notation of this paper. We confirm some of these results but not all, as explained below.

The ``dispersal region" in the phase diagram of \cite{Helfer:2016ljl}, which we understand to be the transition branch of solutions in the equations of motion, disappears above the triple point at $f\gtrsim 0.3\tilde{M}_p$. We observe this as the disappearance of the transition branch, which occurs for $\d \gtrsim 10^{-1}$ (the purple curve in Fig. \ref{fig:MvsR}), roughly confirming their result for $f_{TP}$; we do not, however, confirm the mass at the triple point, which is a factor of few larger in their result compared to ours. It is possible this difference originates in the use of the ``classical" equations of motion in \cite{Helfer:2016ljl} (i.e. a cosine rather than a Bessel function in the self-interaction potential); we have pointed out previous that this change, when evaluated on the transition branch of axion stars, leads to differences compared to the semi-classical analysis we have used here \cite{Eby:2019ntd}.

\begin{figure}[t]
\centering
\includegraphics[scale=0.60]{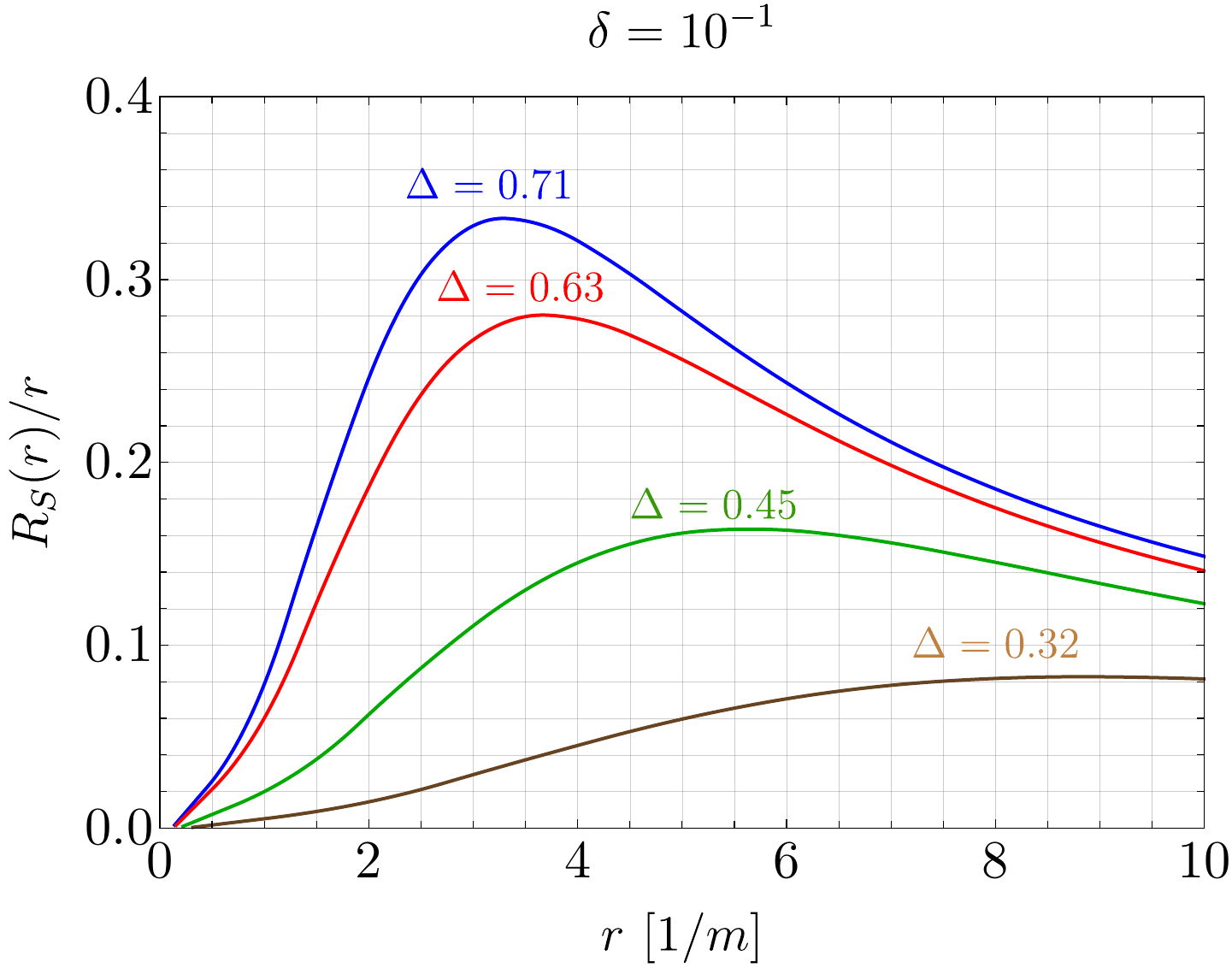}\,
\caption{The ratio of the Schwarzschild radius $R_S(r)$ to the radius in the axion star $r$, for radii less than or equal to the total radius $R_{99}$, at $\d=10^{-1}$ and at different (labeled) values of $\D$. When the ratio is $\ll 1$, the star is far from the General Relativistic limit.}
\label{fig:RSratio}
\end{figure}

Secondly, in spite of analyzing states at large binding energy, we find General Relativistic effects to be negligible everywhere, and therefore do not confirm the formation of black holes observed in \cite{Helfer:2016ljl,Widdicombe:2018oeo}. For each solution at large $\d$, we confirm first that the Schwarzschild radius $R_S = 2\,G\,M$ is always much smaller than the radius of the star $R_{99}$. We further checked that the condition $R_S(r) = 2\,G\,\Mcal(r) < r$ is satisfied inside the star at every $r<R_{99}$, as most of the density is in the inner region. For $\d=0.1$ (near the triple point of \cite{Helfer:2016ljl}), we see in Fig. \ref{fig:RSratio} that the ratio $R_S(r)/r \ll 1$ for all points in our solution space, and therefore these states are not black holes. Note however, that the analyses of \cite{Helfer:2016ljl,Widdicombe:2018oeo} were dynamical, focusing on the collapse of axionic objects, whereas ours is static by construction; the results of this work apply to structurally stable (or metastable) configurations only. In those works the initial profiles were also far from the physical axion star profiles we analyze here. These factors may account for any discrepancy between the two results, though a more thorough investigation may be warranted.

\bibliography{LargefASts}
\bibliographystyle{unsrtnat}

\end{document}